# Effects of Ga$^+$ milling on InGaAsP Quantum Well Laser with mirrors etched by Focused Ion Beam

F. Vallini[1,*], D. S. L Figueira[1,*], P. F. Jarschel[1], L. A. M. Barea[1], A. A. G. Von zuben[1], A. S. Filho[1], and N. C. Frateschi[1,2]

[1] Instituto de Física "Gleb Wataghin," Universidade Estadual de Campinas-UNICAMP, São Paulo, 13083-970, Brazil.

[2] Center for Semiconductor Components, Universidade Estadual de Campinas-UNICAMP, São Paulo, 13083-870, Brazil.

* authors with equal contributions

**Abstract**

InGaAsP/InP quantum wells (QW) ridge waveguide lasers were fabricated for the evaluation of Ga$^+$ Focused Ion Beam (FIB) milling of mirrors. Electrical and optical proprieties were investigated. A 7% increment in threshold current, a 17% reduction in external quantum efficiency and 15 nm blue shift in the emission spectrum were observed after milling as compared to the as cleaved facet result. Annealing in inert atmosphere partially revert these effects resulting in 4% increment in threshold current, 11% reduction in external efficiency and 13 nm blue shift with the as cleaved result. The current-voltage behavior after milling and annealing shows a very small increase in leakage current indicating that optical damage is the main effect of the milling process.

**Indexing terms**: Focusing Ion Beam, Milling, Semiconductor Laser.

**Introduction**

Focused Ion Beam (FIB) has been widely used in Transmission Electronic Microscopy (TEM) and recently has been used for micro and nanofabrication [1-11]. FIB is useful for monolithic fabrication because it does not require lithography and allows nanometric etching resolution [10]. Several works have reported the use of this technique for ion implantation, due to the high beam intensity around 30keV and the great precision in beam positioning [5-8]. Also, FIB has been employed in silicon based devices. The devices being fabricated using this approach are photonic crystals, Silicon dots formation, thin films formation, slot waveguides, micro resonators, etc [1, 3, 11-13]. In III-V compounds the application of FIB is more recent, and good results have been demonstrated with the fabrication of quantum cascade lasers, photonic crystals, microcavities and laser diodes [2, 8, 10, 14].

Although FIB may be very suitable for device fabrication, usually etching leads to deposition and even implantation of residual ions over the substrate that results in alterations of materials optical and electrical properties [15-16]. The effects of ion implantation were demonstrated with the SRIM software, based on Monte Carlo simulations for scattering analysis [16].

In this work, we study the effect of the gallium ions milling in the optical and electrical properties of ridge waveguide lasers with FIB milled facets.

**Experimental Methods**

A cleaved facet ridge waveguide laser was fabricated using an epitaxial structure with InGaAsP based multi-quantum well active and guiding regions sandwiched by InP cladding layers all grown on a n-InP (001) substrate by Metallorganic Chemical Vapor Deposition (MOCVD). *N* and *p* doping was obtained with silicon and zinc, respectively. Highly doped lattice matched $p^+$-InGaAs was used as the top contact layer. The ridge is 6.0 μm wide and 2.7 μm tall. This geometry was obtained using wet etching with $H_2SO_4/H_2O_2/H_2O$ (1:8:40) for the $p^+$ InGaAs layer, and $HCl/H_2O$ (3:1) for the InP cladding layer. Subsequently to etching, the entire surface was covered with a 3000 Å thin film of $Si_3N_4$ deposited with an Electron Cyclotron Resonance (ECR) plasma system. Metallization windows were opened on the top of the ridges using $SF_6$/Ar Reactive Ion Etching (RIE) process. Ti/Pt/Au and Ni/Ge/Au/Ni/Au were deposited by e-beam evaporation to form the p and n contacts, respectively. The samples were cleaved into bars with several lasers with 530 μm cavity length. Electrical and optical characterization was performed on all as-cleaved devices.

After characterization, the same samples were taken to a FEI NOVA 200 Dual Beam System FIB/TEM (Focused Ion Beam/Transmission Electronic Microscope), where their cleaved faces were etched with a total depth of 600 nm each. A 30 keV $Ga^+$ beam with 1 nA emission was employed. Fig. 1 shows a Scanning Electron Microscopy (SEM) micrograph of laser facet (a) as-cleaved and (b) after milling. A good morphology is achieved after milling.

After FIB milling, a second set of electrical and optical characterization was performed. It is known that studies on annealing that can help eliminating defects caused by ion milling have been done. Typical annealing temperatures for this process range from 200 ºC to 600 ºC [16]. However, we must keep the temperature below the contact alloy temperature, 420 °C. We have chosen a 300 °C treatment under $N_2$ atmosphere for a period of 1 hour. Finally, a third set of electrical and optical measurements was performed.

**Results and Discussion**

Fig. 2 shows the plot of the second derivative of light power output versus injection current ($d^2L/dI^2$) for the devices after cleaving, after cleaving plus milling and after cleaving, milling and annealing. The inset shows the light output power versus current plot (LxI) for the same devices. Threshold current increased from 21.4 mA to 23.0 mA after milling, reducing finally to 22.3 mA after the annealing. Threshold current was defined as the current corresponding to the maximum $d^2L/dI^2$ value. The same behavior is also observed for the external quantum efficiency. After milling, the external efficiency reduced by 17 % recovering 6 % after the annealing. Fig. 3 shows the emission spectrum of the laser in the three stages of the process. After milling the emission peak blue shifted 15 nm. The annealing reduces the blue shift by 2 nm. Fig. 4 shows the current-voltage curve (IxV) for the devices as cleaved, after milling and annealing and after only annealing for comparison. Annealing increases the dark and reverse current by 2 nA. Essentially the electrical differences between devices with and without milling are negligible.

Since milling or milling followed by annealing increase leakage current by less than 1 nA, the approximately 1 mA increase in threshold is entirely caused by mirror losses. The blue shift also corroborate with this hypothesis since it is an indication of an increase in threshold carrier density and the band filling effect. If threshold increase was caused by leakage current we would expect a red shift caused by Joule heating [17]. Also, Monte Carlo simulations indicate that at incidence angle of 90° the implantation of $Ga^+$ ions is approximately 50 nm inside the longitudinal direction on each face of the laser

[16], reinforcing the idea that the effects over the junction are minimal. The annealing process partially revert these defects, causing a reduction in threshold current from the pre-annealing state.

Assuming the damage is causing only variation in mirror losses, the reflectivity variation can be estimated. We have employed the Hakki-Paoli method [19] for a cleaved facet lasers fabricated with the same epitaxial material and the net modal gain, *g*, dependence with injected current, *I*, obtained is $g(I) = -0.176I^2 + 7.304I - 54.114$. With the gain, we calculate a reflectivity of 26.2 % for the cleaved facet. This result is in good agreement with the expected mirror reflectivity obtained from the Fresnel expression. The increase in threshold current after the milling and annealing can be used to estimate the reduction in facet reflectivity. We obtained a 23.4 % reflectivity after milling and a 24.6 % after milling with subsequent annealing. Therefore, the milling process plus annealing treatment provides a very high quality mirror with only 2 % deterioration.

**Conclusions**

We demonstrated the influence of $Ga^+$ ion milling on InGaAsP laser facets. Very high quality mirrors were obtained after milling and annealing with only a 2 % reduction in reflectivity and no electrical degradation. Annealing after milling is essential to reduce the damages caused by the $Ga^+$ ions. Giving the high quality of the process, we expect it to be very suitable for the fabrication of micro-cavity structures that demand exceptionally high reflectivity.

**Acknowledgments:** The authors would like to thank Antônio Augusto Von Zuben for the help on the laser fabrication. This work was supported by the Coordenação de Aperfeiçoamento de Pessoal de Ensino Superior (CAPES), Conselho Nacional de Desenvolvimento Científico e Tecnológico (CNPq), the Fundação de Amparo à Pesquisa do Estado de São Paulo (FAPESP) and the Centro de Pesquisa em Óptica e Fotônica (CEPOF).

Figure Captions

Figure 1: Scanning Electron Microscopy (SEM) micrograph of laser facet (a) as-cleaved and (b) after milling with a FIB.

Figure 2: Second derivative of light power output versus injection current ($d^2L/dI^2$) for the devices as cleaved (dotted line), after cleaving and FIB milling (dashed line), and after cleaving, milling and annealing (solid line). The inset shows the light output power versus current plot (LxI) for the same devices.

Figure 3: Emission spectrum of the laser in the three stages of the process, as cleaved (dotted line), after FIB milling (dashed line), and after milling and subsequent annealing (solid line).

Figure 4: Current-voltage curves obtained in three different conditions, as cleaved (dotted line), after FIB milling (dashed line), and after annealing only (solid line).

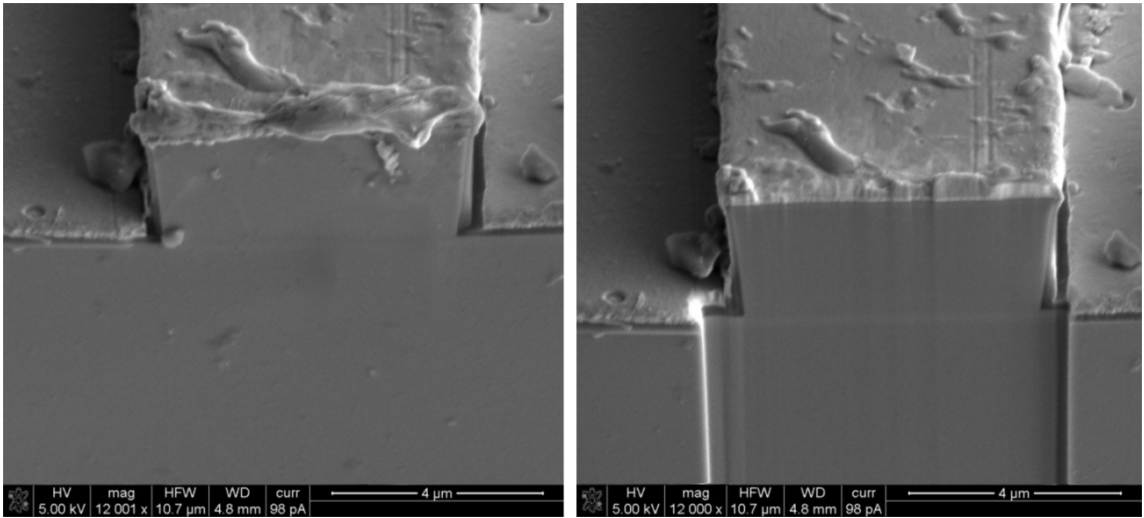

Figure 1.

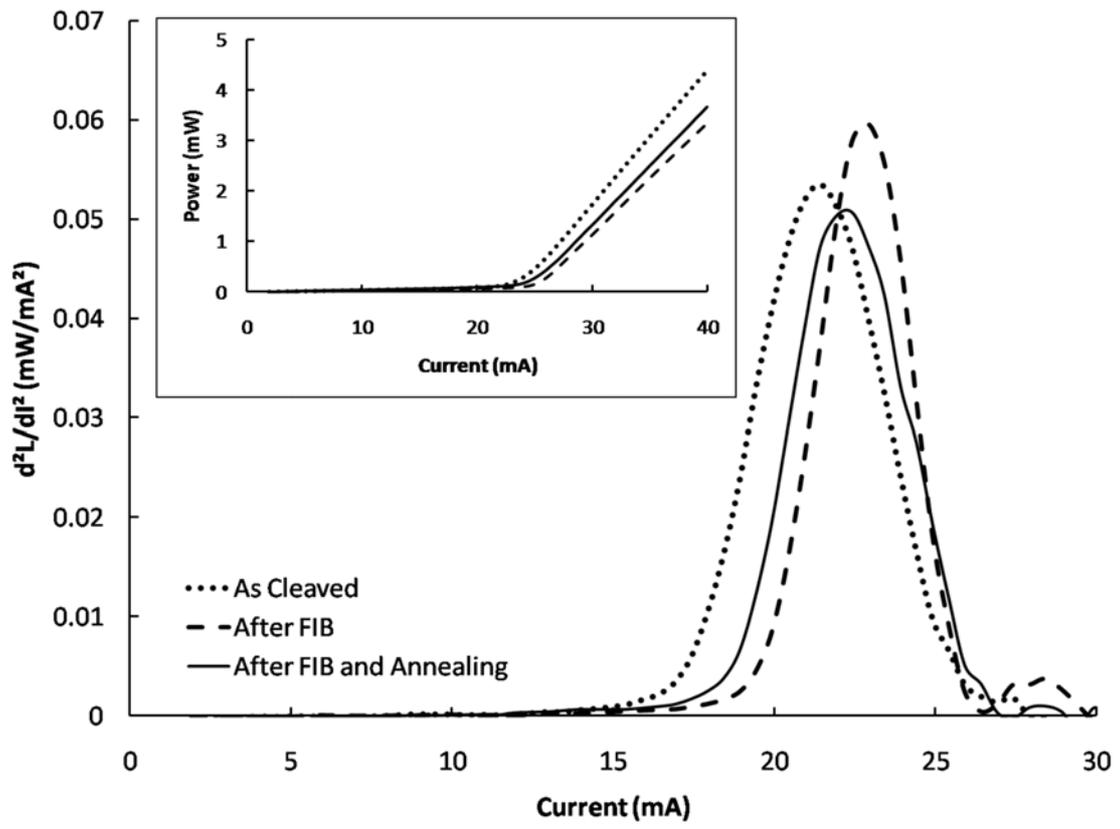

Figure 2.

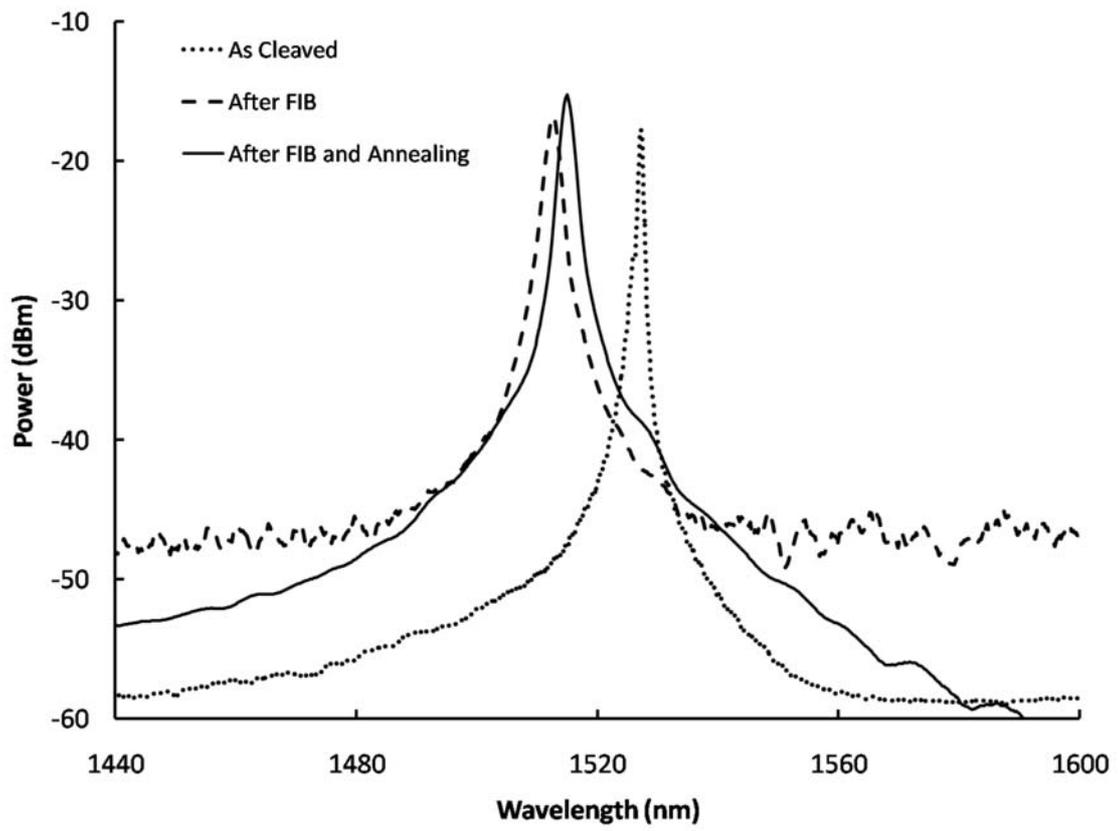

Figure 3.

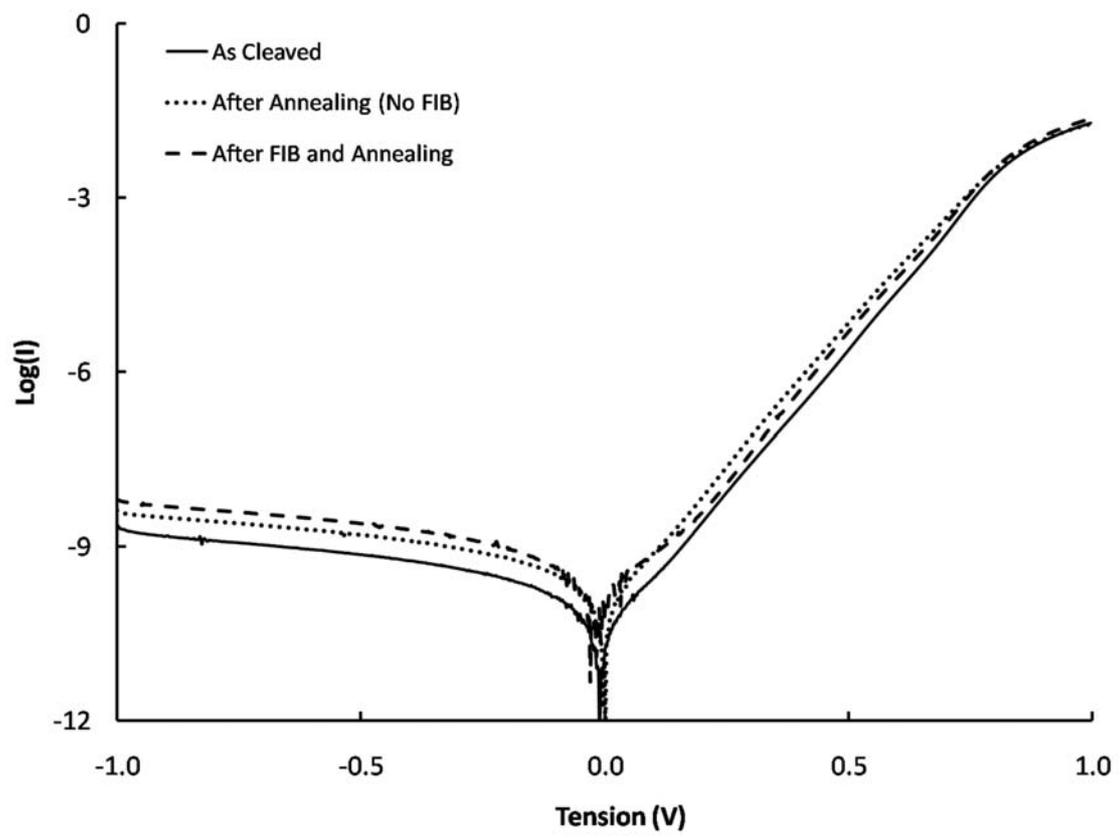

Figure 4.